\documentclass[11pt,a4paper]{article}

\usepackage{amsfonts}
\usepackage[dvips]{graphicx}
\usepackage{wrapfig}
\usepackage{amsthm}
\usepackage{epsfig}
\usepackage{t1enc}
\usepackage{amssymb}
\usepackage{latexsym}
\usepackage{bm}

\title{\bf Group velocity of gravitational waves
\\
in an expanding universe}
\author{Vladim\'\i r Balek\footnote{e-mail address: balek@fmph.uniba.sk}\ \ and
Vratko Pol\' ak\footnote{e-mail address: polak@fmph.uniba.sk}
\\
{\it Department of Theoretical Physics,
Comenius University, Mlynsk\'a dolina}
\\
{\it 842 48 Bratislava, Slovakia}}

\begin{document}
\maketitle
\maketitle\abstract

{The group velocity of gravitational waves in a flat
Friedman-Robertson-Walker universe is investigated. For plane
waves with wavelength well inside the horizon, and a universe
filled with an ideal fluid with the pressure to density ratio less
than 1/3, the group velocity is greater than the velocity of
light. As a result, a planar pulse of gravitational waves
propagating through the universe during the matter/dark energy
dominated era arrives to the observer with the peak shifted
towards the forefront. For gravitational waves emitted by
inspiralling supermassive black holes at the edge of the
observable universe, the typical shift that remains after the
effects of nonplanarity are suppressed is of order of ten
picoseconds.}


\section{Introduction}
Recent efforts in the construction of large detectors of
gravitational waves based on laser interferometry (for a review,
see \cite{roho}) revived interest in investigations of emission,
propagation and detection of gravitational waves. Topics that have
been studied in some detail in the past include propagation of
weak gravitational waves in a Friedman-Robertson-Walker (FRW)
universe \cite{{cal}, {sofa}, {noon}, {mawy}}. The main result is
that the waves do not obey Huygens' principle unless the universe
is filled with pure radiation (so that its scalar curvature is
zero). Otherwise there appears a 'tail' in the Green's function,
coming from the backscattering of waves on the spacetime
curvature. This contrasts with the behavior of the electromagnetic
waves, whose 'tail', if present, consists of pure gauge.

\vskip 2mm A kinematic characteristic often used in the
description of wave propagation is the group velocity. Strictly
speaking, it is defined for modulated waves only, as the velocity
with which 'wave groups' travel through space. However, one
commonly speaks of the group velocity of wave pulses of limited
duration, too, referring to the velocity of the peak of the
envelope of the pulse. (For compact pulses, this can be identified
with the signal velocity, defined as the velocity with which the
main part of the signal arrives to the observer \cite{bri}.) Here
we address the question what is the group velocity of
gravitational waves in a flat FRW universe. We restrict ourselves
to a flat 3-geometry because it is easier to deal with than a
closed or open one, and is preferred by observations. From the
results on the validity of Huygens' principle it follows that if
the universe is filled with pure radiation, plane gravitational
waves advance as a whole with the velocity of light (along null
geodesics with respect to the background metric). In this case the
group velocity is, of course, also the velocity of light. For
other kinds of cosmic environment the group velocity is presumably
different; moreover, while the 'tail' of the Green's function lays
entirely inside the light cone, there is no {\it a priori} reason
for the 4-vector of the group velocity to lay inside it, too. A
superluminal group velocity does not imply violation of causality
since the information carried by a pulse arrives to the observer
with the forefront, which travels with the velocity of light (as
follows from the Hadamard discontinuity formalism, see
\cite{had}). This is similar to what happens when light passes
through a medium with anomalous dispersion. As demonstrated
experimentally, a pulse of light can propagate, without being
significantly distorted, even with negative group velocity
\cite{exp}. However, no problem with causality arises since in a
dispersive medium, too, the forefront of a pulse travels with the
velocity of light \cite{bri}.

\vskip 2mm In section \ref{sec:vg} we compute the group velocity
of gravitational waves using a well-known formula from optics, in
section \ref{sec:vg1} we make sure that the formula can be applied
to the problem under study, and investigate its subtleties, within
the theory of wave propagation in a time-dependent medium, and in
section \ref{sec:concl} we discuss the results. We use the
signature of metric tensor $(-, +, +, +)$ and a system of units in
which $c = G = 1$.


\section{Group velocity: calculation}
\label{sec:vg}

Consider weak gravitational waves propagating in a flat FRW
universe. The background metric is
$$ds^2 = a^2 \left(- dt^2 + d{\bf x}^2\right),$$
where $t$ is conformal time, $x^i$ are comoving coordinates and $a
= a(t)$ is the scale parameter. Gravitational waves are described
by the perturbation to the metric of the form
$$a^2 e_{ij} dx^i dx^j$$
where $e_{ij}$ is the traceless transversal part of the total
perturbation, $e_{ii} = e_{ij,j} = 0$. Denote the derivative with
respect to $t$ by an overdote. The linearized Einstein equations
without anisotropic stress yield (see, for example, \cite{ll})
$$\ddot e_{ij} + 2H \dot e_{ij} - \triangle e_{ij} = 0,$$
where $H = \dot a/a$ is the Hubble parameter. After passing from
$e_{ij}$ to $E_{ij} = a e_{ij}$, one arrives at an equation in
which the damping term is traded for a term with an external
field. The equation reads
\begin{equation}
\ddot E_{ij} - \triangle E_{ij} + V E_{ij} = 0, \label{eq:E}
\end{equation}
where
\begin{equation}
V = - \dot H - H^2. \label{eq:V}
\end{equation}
Suppose the universe is filled with a perfect fluid (possibly,
including dark energy) and denote the density of the fluid by
$\rho$ and the pressure of the fluid by $p$. Then,
$$H^2 = \frac{8\pi}3 \rho a^2,\ \ \dot H = - \frac{4\pi}3
(\rho + 3p)a^2,$$ so that
\begin{equation}
V = - \frac{4\pi}3 (\rho - 3p). \label{eq:V1}
\end{equation}
For a mixture of nonrelativistic matter ($p \ll \rho$), radiation
($p = \rho/3$) and dark energy ($p = -\rho$) one obtains $V \le
0$, with the equality valid only if the universe is filled with
pure radiation. For $E_{ij}$, we have a wave equation in {\it
flat} spacetime, with a time-dependent dispersion relation
\begin{equation}
\omega = \sqrt{k^2 + V}, \label{eq:disp}
\end{equation}
where $\omega$ is frequency and $k$ is wave number (the magnitude
of the wave vector). Suppose $\omega$ is real, that is, there
holds either $V = 0$ or $V < 0$ and $k > |V|$ (the wavelength is
less than the Jeans length). Then, the field $E_{ij}$ is in an
oscillatory regime and one can ask how fast does it propagate.
According to the textbook formula (see, for example, \cite{born}),
the group velocity in a medium with an isotropic dispersion
relation is $v = \partial_k \omega$. This yields
\begin{equation}
v = \frac k{\sqrt{k^2 + V}}, \label{eq:v}
\end{equation}
so that from $V \le 0$ we have $v \ge 1$. Note that $v =
1/v_{ph}$, where $v_{ph} = \omega/k$ is the phase velocity.

\vskip 2mm When computing $v$, we have ignored the fact that
$\omega$ depends on $t$. However, the formula $v = \partial_k
\omega$, supposed to be valid for $\omega = \omega(k)$, should
stay valid also for $\omega = \omega(k, t)$ if the period of the
oscillations of the field is much less than the time within which
$\omega$ varies considerably. The period is $2\pi/\omega$ and the
time scale of the variation of $\omega$ is $\omega/|\dot \omega|$,
therefore (\ref{eq:v}) must be supplemented by the condition
\begin{equation}
\frac {|\dot \omega|}{\omega^2} \ll 1. \label{eq:cond}
\end{equation}
After inserting here from (\ref{eq:disp}) we find
$$k^2 \gg |V| + |\dot V|^{2/3}.$$
The universe is dominated first by radiation, then by matter, and
then by dark energy, with the function $a(t)$ subsequently of the
form $a \propto t$, $(t - t_0)^2$ and $(t_1 - t)^{-1}$. When using
these functions to compute $V(t)$, we find that in the first era,
as we already know, $V = 0$, and in the next two eras $V \sim -
\Delta t^{-2}$, where $\Delta t = t - t_0$ and $t_1 - t$ in the
second and third era respectively. The time $\Delta t$
approximately equals the {\it horizon length}, if 'horizon' is
understood as particle horizon in the matter dominated era and
event horizon in the dark energy dominated era. For $V \sim -
\Delta t^{-2}$ there holds $\dot V \sim \pm \Delta t^{-3}$, so
that the two terms in the condition for $k$ are of the same order
of magnitude. As a result, the condition simplifies to
\begin{equation}
k^2 \gg |V|. \label{eq:cond1}
\end{equation}
From $V \sim - \Delta t^{-2}$ we find $k \gg \Delta t^{-1} \sim$
(horizon length)$^{-1}$; thus, in order that a wave packet
propagates with the group velocity (\ref{eq:v}), its mean
wavelength must lay well within the horizon. For such packets, the
group velocity is close to 1, with the correction of the form
\begin{equation}
\Delta v = - \frac V{2k^2}. \label{eq:Dv}
\end{equation}

\vskip 2mm Light signals (particles moving along null geodesics
with respect to the background metric) have coordinate velocity 1.
Thus, the constraint $v \ge 1$ means that the gravitational waves
either propagate with the velocity of light or are superluminal.
The former case occurs in the radiation dominated era and the
latter in the matter/dark energy dominated era. When combining
equations (\ref{eq:V1}) and (\ref{eq:Dv}), we find that the group
velocity differs from the velocity of light by
$$\Delta v = \frac{4\pi}3 (\rho - 3p)\varkappa^{-2},$$
where $\varkappa = k/a$ is the true wave number. For an estimate,
suppose the universe is filled with incoherent dust ($p = 0$).
Then $\rho \sim \tau^{-2}$, where $\tau$ is cosmological time, and
the correction to $v$ is of order $(\varkappa \tau)^{-2}$. As a
source of gravitational waves, consider a binary system consisting
of two black holes with masses $M \sim 10^6 M_\bigodot$, orbiting
around each other with the period $T \sim 10^4s$. (The period
drops to this value about a month before the merger.) Suppose,
furthermore, that the system is at the edge of the observable
universe, at the distance $l \sim 3$ Gpc from Earth. The time it
takes for a packet of gravitational waves to propagate from the
binary to Earth is $\tau_{prop} \sim \tau \sim 10$ Gy, and the
time by which the peak of the packet arrives earlier due to the
correction to $v$ is
$$\Delta \tau \sim (\varkappa \tau)^{-2} \tau_{prop} \sim \frac
{T^2}{(2\pi)^2\tau} \sim 10^{-11} \mbox{s}.$$ For a universe with
the actual values of cosmological parameters, $H_0 = 71$ km
s$^{-1}$ Mpc$^{-1}$, $\Omega_m = 0.26$ and $\Omega_\Lambda =
0.74$, and a source with the period $T = 1$ hour at the redshift
of the most distant quasar, $z = 6.4$, a detailed calculation
yields $\Delta \tau = 6.6 \times 10^{-11}$ s.


\section{Group velocity: theory}
\label{sec:vg1}

Consider a field $u$ satisfying the same equation as $E_{ij}$,
\begin{equation}
\ddot u  - \triangle u + Vu = 0. \label{eq:u}
\end{equation}
This can describe gravitational waves in an expanding universe as
well as light in a homogeneous time-dependent optical medium.
Thus, the following considerations are only loosely related to
general relativity; instead, one can view them as a discussion of
a special problem of optics.

\subsection{Waves with definite wavelength}

Choose $u$ in the form of a plane wave with a definite wavelength
propagating along the $x$-axis, $u = U(t) e^{ikx}$. The wave
number $k$ can now assume any real value, however, for our
purposes it is sufficient to suppose as before $k > 0$. From
(\ref{eq:u}) we obtain
\begin{equation}
\ddot U + \omega^2 U = 0, \label{eq:U}
\end{equation}
with the function $\omega = \omega(k, t)$ defined in
(\ref{eq:disp}). If we exchange $t$ with $x$ and $\omega$ with
$k$, we obtain an equation for monochromatic light in a static
inhomogeneous medium; and if we replace, in addition, $\omega$ by
$E$ (the energy of the particle) and $k$ by $p$ (the momentum of
the particle), we obtain Schr\" odinger equation in one dimension
with a modified relation between $p$ and $E$, $p = \sqrt{E^2 + V}$
instead of $p = \sqrt{2(E - V)}$.

\vskip 2mm For light propagating in a medium with dispersion
relation $\omega = \omega(k, t)$, the high-frequency condition
(\ref{eq:cond}) defines the geometrical optics approximation. An
analogical condition is known to define quasiclassical, or WKB,
approximation in quantum mechanics \cite{ll3}, and geometrical
optics approximation in a static inhomogeneous medium in optics
\cite{ll8}. To obtain an approximate expression for $U$, write $U
= e^{- i\psi}$ and expand the complex phase $\psi$ in the powers
of $\omega^{-1}$. If we denote the terms of the order minus one
and zero by $\psi_0$ and the terms of the first order by $\Delta
\psi$, we have (see the corresponding formulas in \cite{ll3})
\begin{equation}
\psi_0 = \int \omega dt - \frac i2 \log \omega,\ \ \Delta \psi = -
\frac {\dot \omega}{4\omega^2} - \frac 18 \int \frac {\dot
\omega^2} {\omega^3}dt. \label{eq:psi}
\end{equation}
The signs in these expressions are chosen in such a way that the
wave propagates in the positive $x$-direction. The first
expression determines the oscillating field in the geometrical
optics approximation, and the second expression can be included
into the slowly varying amplitude and regarded as the
post-geometrical optics correction to it.

\vskip 2mm If the dispersion relation assumes the form
(\ref{eq:disp}), the condition (\ref{eq:cond}) for $\omega$
implies the condition (\ref{eq:cond1}) for $k$. (This holds
irrespective of the form of $V(t)$. In general, we have also
condition $k^2 \gg |\dot V|^{2/3}$, however, it can be skipped if
we choose $V \sim - \Delta t^{-2}$.) With $k$'s satisfying the
condition (\ref{eq:cond1}), the dispersion relation can be written
approximately as
\begin{equation}
\omega = k + \frac V{2k}. \label{eq:disp1}
\end{equation}
The fact that we have replaced the condition for $\omega$ by the
condition for $k$ suggests that the concept of geometrical optics
approximation must be reconsidered. The small parameter of the
theory is $(k\Delta t)^{-1}$, where $\Delta t$ is the time scale
of the variation of $V$, rather than $(\omega \delta t)^{-1}$,
where $\delta t$ is the time scale of the variation of $\omega$,
therefore it is advisable to expand all quantities in the powers
of $k^{-1}$ rather than $\omega^{-1}$. In particular, the
approximate expression for $\omega$ should be written $\omega =
\omega_0 + \Delta \omega$, with the frequency in the geometrical
optics approximation $\omega_0 = k$ and the post-geometrical
optics correction to it $\Delta \omega = V/(2k)$. This yields
\begin{equation}
\psi_0 = kt - \frac i2 \log k,\ \ \Delta \psi = \frac 1{2k}\int V
dt. \label{eq:psi1}
\end{equation}
The previously obtained expression for $\Delta \psi$ does not
contribute here since it is of order $k^{-3}$. Note, however, that
it {\it would} contribute if we used cosmological time instead of
conformal, since then it would be of order $k^{-1}$.

\subsection{Group velocity of a Gaussian packet with infinitesimal dispersion}

From plane waves with a definite wavelength one can form a planar
wave packet
\begin{equation}
u(x, t) = \int \limits_{-\infty}^\infty a (K) e^{-i\psi(K, t) +
iKx} dK, \label{eq:pacx}
\end{equation}
where $a$ is an overall amplitude peaked at the given wave number
$k$ and differing significantly from zero only in an interval of a
size $\sigma \ll k$. Note that the integral contains modes with
all $k$'s, and we have derived formulas for $\psi (k, t)$ only for
$k$ positive and large enough to satisfy the high-frequency
condition. However, the modes with $k$ either positive and small
or negative do not have significant effect on the results in the
limit we are interested in, therefore we can describe them by an
extrapolated formula for large positive $k$.

\vskip 2mm Define the envelope of the wave packet as the curve
$|u(x, t)|$ at fixed $t$, and the group velocity $v$ as the
velocity with which the peak of the envelope moves along the
$x$-axis. That is,
\begin{equation} v = \dot x_M, \label{eq:dfvg}
\end{equation}
where $x_M$ is the value of $x$ at which the function $|u(x, t)|$
has maximum at fixed $t$. To obtain an approximate expression for
$v$, let us adopt two simplifications. First, consider a Gaussian
packet,
\begin{equation}
a_G = \frac 1{\sqrt{2\pi}\sigma } \exp \left(-\frac {\Delta
K^2}{2\sigma^2}\right), \label{eq:gauss}
\end{equation}
and second, expand the phase in the powers of $\Delta K = K - k$,
\begin{equation}
\psi(K, t) = \psi + \xi \Delta K + \frac 12 \eta \Delta K^2 +\frac
16 \zeta \Delta K^3 + \ldots, \label{eq:exppsi}
\end{equation}
where $\psi = \psi(k, t)$, $\xi = \partial_k \psi$, $\eta =
\partial^2_k \psi$, $\zeta = \partial^3_k \psi $, $\ldots$, and
compute $v$ in the linear approximation in which the expansion is
truncated after the $\xi$-term.

\vskip 2mm Let $\cal U$ be the function obtained by factorizing
out the quickly oscillating phase factor from $u$, $u = e^{-i\psi
+ ikx}\cal U$. For a Gaussian packet, $\cal U$ in the linear
approximation is
$${\cal U}_{lin} = e^{-\frac 12 \sigma^2 (x - \xi)^2},$$ and its
magnitude is
$$|{\cal U}_{lin}| = e^{- \frac 12 \sigma^2 [(x - \xi_R)^2
- \xi_I^2]},$$ where the indices $R$ and $I$ denote the real and
imaginary part of the quantity in question. We can see that the
wave packet in $x$-space is Gaussian, like in $k$-space, with the
width $\sigma_x = 1/\sigma$ and the peak located at
\begin{equation} x_M = \xi_R. \label{eq:xlin}
\end{equation}
For a general time-dependent medium we have
$$\xi_{0R} = \int \partial_k \omega dt,\ \ \Delta \xi_R =
- \frac 14 \partial_k \left(\frac {\dot \omega}{\omega^2} + \frac
12 \int \frac {\dot \omega^2} {\omega^3} dt\right).$$ From the
first expression we obtain the leading term in $v$, which is just
the standard group velocity cited in the previous section,
\begin{equation}
v_0 = \partial_k \omega. \label{eq:v0}
\end{equation}
The second expression yields the correction
\begin{equation} \Delta v = - \frac 14 \partial_k \left(\frac
{\ddot \omega}{\omega^2} - \frac {3\dot \omega^2} {2\omega^3}
\right). \label{eq:Dv1}
\end{equation}
For a medium with the dispersion relation (\ref{eq:disp}) we have
$$\xi_{0R} = t,\ \ \Delta \xi_R = - \frac 1{2k^2} \int V dt,$$
which yields $v = 1 + \Delta v$ with $\Delta v$ given by equation
(\ref{eq:Dv}). In this case the group velocity is given, within
the current accuracy, entirely by the formula $v =
\partial_k \omega$.

\vskip 2mm If we include higher order terms into the $\Delta
K$-expansion of $\psi$ into the calculation, we obtain corrections
of order $\sigma^2$, $\sigma^4$, $\ldots$ to $\cal U$ near $x =
x_M$, producing corrections of order $\sigma^2$, $\sigma^4$,
$\ldots$ to $x_M$. Thus, the values of $x_M$ and $v$ obtained for
a Gaussian packet in the linear approximation can be viewed as
exact values in the limit $\sigma = 0$ (vanishing width of the
packet in $k$-space). Equivalently, one can speak of the limit
$\sigma_x = \infty$ (infinite width of the packet in $x$-space).

\subsection{Effects of finite dispersion and nongaussianity}

Consider first a Gaussian packet beyond the linear approximation.
To find the first correction to the expression for $x_M$ obtained
before may seem trivial, because if we include the $\eta$-term
into $\psi(K, t)$, we arrive at an integral that can still be
calculated explicitly. However, as shown in the appendix, the
correction contains also a contribution of the $\zeta$-term which
is comparable with the contribution of the $\eta$-term or greater.
After taking into account this contribution we find that (see
equation (\ref{eq:dxgauss1}) and the comment preceding equation
(\ref{eq:gaussest}))
$$\mbox{the first correction to}\ x_M \sim \frac 1\kappa
\eta_R \sigma^2,$$ where $\kappa$ is a typical interval of $k$
within which $\omega$ varies, $\kappa = \omega/|\partial_k
\omega|$. If $\Delta v$ is to stay the leading correction to
$v_0$, the expression on the r.h.s. must be much smaller in the
absolute value than the correction to $x_M$ from which $\Delta v$
has been computed,
\begin{equation} \frac 1\kappa |\eta_R| \sigma^2 \ll |\Delta
\xi_R|. \label{eq:gausscond}
\end{equation}
Of course, the condition is relevant only if the effect of higher
corrections to $x_M$ coming from finite $\sigma$ is negligible;
that is to say, if the perturbation theory with small parameter
$\sigma$ can be employed. The corresponding conditions are
summarized in equation (\ref{eq:ptcond}). It turns out that
$\sigma$ is small enough if it is much less, at the same time,
than $1/|\xi_I|$, $1/\sqrt{|\eta_R|}$ and $\kappa$.

For nongaussian packets, write the linear part of $\cal U$ as
$${\cal U}_{lin} = \int \limits_{-\infty}^\infty b(K, t)
e^{i\Delta K X} dK,$$ where $b (K, t)= a e^{\xi_I \Delta K}$ and
$X = x - \xi_R$. The additional correction to $x_M$, supplementing
that from which $\Delta v$ has been computed, equals the value of
$X$ at which the function $|{\cal U} (X, t)|$ has maximum. Denote
this value by $X_M$. Since $b$ depends on $t$, $X_M$ is expected
to depend on $t$, too; and since $t$ appears in $b$ through
$\xi_I$, which has the same physical dimension as $X_M$, the
leading time-dependent term in $X_M$ is expected to be
proportional to $\xi_I$ with a coefficient of order 1. Thus, if
one considers a generic nongaussian packet and wants $\Delta v$ to
be the dominant correction to $v$, one must have, in addition to
small $\sigma$, also small $|\xi_I|$. A term proportional to
$\xi_I$ is present in $X_M$, for example, if the packet has
amplitude $(1 + \alpha \Delta K/\sigma)a_G$ with $\alpha = const$.
Such packet has two maxima placed symmetrically with respect to
the center, located at the points
\begin{equation} X_M = \pm \left[const + \frac 1\alpha \xi_I + O(\sigma)\right].
\label{eq:specXm}
\end{equation}
On the other hand, it is easy to construct a wide class of packets
for which the time-dependent part of $X_M$ is {\it exactly} zero
in the linear approximation. For that purpose, we can use the
argument of \cite {born} with exchanged space and time dimensions.
Suppose $a$ equals a real nonnegative function modulo a phase
factor with the phase proportional to $\Delta K$, $a = A
e^{iq\Delta K}$ with $A \ge 0$ and $q = const$. Then
$$|{\cal U}_{lin}|^2 = \int dK dK' B(K, t)B(K', t)\cos[(K - K')
(X + q)],$$ where $B (K, t)= A e^{\xi_I \Delta K}$, and this has
obviously maximum at $X_M = -q$.

A general analysis of nongaussian packets can be found in the
appendix. The estimate of the time-dependent part of $X_M$ in
equation (\ref{eq:genest}) implies that the conditions under which
$\Delta v$ dominates other corrections to $v$ are
\begin{equation} |\xi_I|\ \mbox{and}\ |\eta_R| \sigma
\ll |\Delta \xi_R|. \label{eq:gencond}
\end{equation}
In addition to that, one must again satisfy conditions
(\ref{eq:ptcond}) in order that the perturbation theory with small
parameter $\sigma$ is applicable.

\vskip 2mm In a general time-dependent medium, the quantities
entering the constraints (\ref{eq:ptcond}), (\ref{eq:gausscond})
and (\ref{eq:gencond}) are of order
\begin{equation}
\Delta \xi_R \sim \frac 1{\kappa \omega \delta t},\ \ \xi_I \sim
\frac 1\kappa,\ \ \eta_R \sim \frac {\omega \delta t} {\kappa^2}.
\label{eq:gdrest}
\end{equation}
From these estimates we find that the perturbation theory is
justified only if $\sigma \ll (\omega \delta t)^{-1/2}\kappa$, and
$\Delta v$ cannot be the leading correction to $v_0$ for a general
packet. However, it becomes the leading correction if the packet
is Gaussian and has $\sigma \ll (\omega \delta t)^{-2}\kappa$.
Note that an analogical condition for a quantum mechanical
particle in an external potential, obtained by flipping over space
and time directions and identifying wave characteristics with
mechanical ones, yields a packet that is much larger than the
typical distance on which the potential varies. Thus, the post-WKB
correction to $v$ has no relevance for the description of the
motion of a quantum mechanical particle in an external potential.

\vskip 2mm For a medium with dispersion relation (\ref{eq:disp}),
we must modify the theory slightly in order that the results of
the appendix remain valid. Namely, we must define the phase $\psi$
in the geometrical optics approximation in a different, and
perhaps more natural, way, with omitted $\log k$. This yields an
expression for $U$ stripped of the overall factor $1/\sqrt{k}$,
which can be regarded as absorbed into $a$. Using this 'reduced'
definition of $\psi$, and assuming $V \sim - \Delta t^{-2}$, we
find that the relevant quantities are of order
\begin{equation}
\Delta \xi_R \sim k^{-2} \Delta t^{-1},\ \ \xi_I \sim k^{-3}
\Delta t^{-2}, \ \ \eta_R \sim k^{-3} \Delta t^{-1},
\label{eq:polest}
\end{equation}
(The second estimate comes from $\psi_I = -V/(4k^2)$.) If we also
take into account that $\kappa = k$, we conclude that both sets of
constraints (\ref{eq:ptcond}) and (\ref{eq:gencond}) are satisfied
provided $\sigma \ll k$, which we have assumed from the very
start. Thus, for the 'cosmological' dispersion relation, the
quantity $\Delta v$ coming from the post-geometrical optics
contribution to $\psi$ is the dominant correction to $v$ for all
wave packets whose width in $k$-space is small in comparison with
the typical value of $k$.

\subsection{Exact solution}

\begin{figure}[ht]
\centerline{\includegraphics[width=7cm]{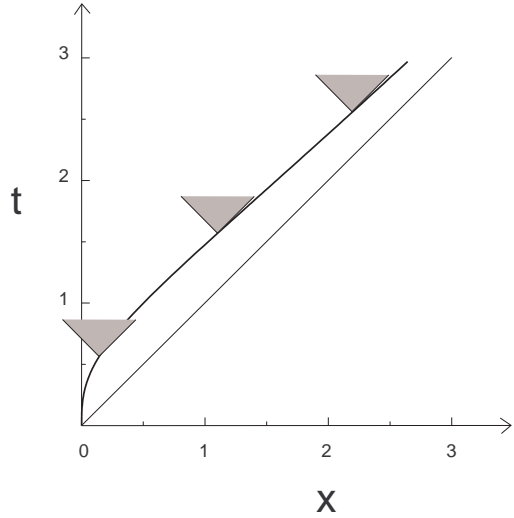}} \caption{\small
Gravitational wave packet in a dust universe} \label{fig:vg}
\end{figure}

To support our conclusions, let us construct an exact solution to
equation (\ref{eq:u}). Suppose
\begin{equation}
V = - \frac 2{t^2}. \label{eq:specV}
\end{equation}
As seen from (\ref{eq:V}), $V$ assumes this form for gravitational
waves in a dust universe (which expands after the law $a \propto
t^2$). First we find solutions with definite wavelength. If we
pass from $U$ to $U/t$ and from $t$ to $kt$, (\ref{eq:U})
transforms into the equation for spherical Bessel function with $l
= 1$. Moreover, we can see from (\ref{eq:psi1}) with omitted
logarithm that the asymptotics of $U$ for $kt \gg 1$ is $U =
e^{-i[kt + 1/(kt)]}$. This yields
$$U = - kt h_1^{(2)} (kt) = \left(1 - \frac i{kt} \right)
e^{-ikt}.$$ Next we form a wave packet with the amplitude $K
a_G/k$. The function $\cal U$ is
$${\cal U} = [1 + i(p - qs)] e^{-\frac 12 s^2},$$
where $s = \sigma (x - t)$, $p = 1/(kt)$ and $q = \sigma/k$. After
inserting this into the condition $\partial_s |{\cal U}|^2 (s_M) =
0$ we find
\begin{equation}
\left[1 - q^2 + (p - qs_M)^2\right]s_M + pq = 0. \label{eq:sM}
\end{equation}
By definition $s_M = \sigma (x_M - t)$, therefore the equation for
$s_M$ determines the worldline of the peak of the packet $x = x_M
(t)$. In fig. \ref{fig:vg} this worldline is depicted for $k = 1$
and $\sigma = 1/3$.
At the beginning, the peak of the packet stays within the light
cone, but then it moves out. This is consistent with the sign
$\Delta v$ assumes at late times according to (\ref{eq:Dv}).

\subsection{Spherical packet}

\begin{figure}[ht]
\centerline{\includegraphics[width=8cm]{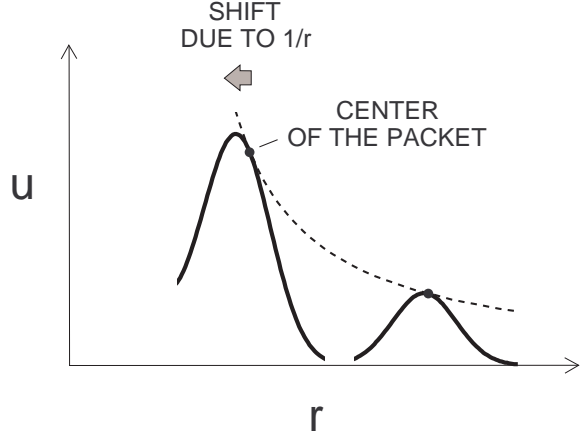}} \caption{\small
Superluminal velocity of a spherical wave packet} \label{fig:spp}
\end{figure}

Consider equation (\ref{eq:u}) with the source $4\pi j$. If we
ignore the field $V$, the solution is
\begin{equation} u^{(0)} ({\bf r}, t) = \int \frac 1R j({\bf
r}', t - R) d^3x', \label{eq:u0}
\end{equation}
where $R = |{\bf r} - {\bf r}'|$. Suppose the source is
nonrelativistic. In the wave zone, $u^{(0)}$ can be written as
$$u^{(0)} \doteq \frac 1r (f + {\bf n}\ .\ \dot {\bf g}) + \frac 1{r^2}
{\bf n}\ .\ {\bf g},\ \ f = \int j({\bf r}', \tau) d^3x',\ \ {\bf
g} = \int j({\bf r}', \tau) {\bf r}' d^3x',$$ where $\tau = t - r$
and ${\bf n} = {\bf r}/r$. The magnitude of $u^{(0)}$ is
$$|u^{(0)}| \doteq \frac 1r (|f| + {\bf n}\ .\ {\bf G}_1) + \frac 1{r^2}
{\bf n}\ .\ {\bf G},\ \ {\bf G} = \frac 1{|f|} \mbox{Re} (f {\bf
g}^*),\ \ {\bf G}_1 = \frac 1{|f|} \mbox{Re} (f \dot {\bf g}^*).$$
We are interested in the radial coordinate $r_M^{(0)}$ at which
$|u^{(0)}|$ has maximum for the given direction of the propagation
of the packet $\bf n$. To compute this quantity, we expand the
derivative
$$\partial_r |u^{(0)}| \doteq - \frac 1r \left(|f|\dot{\ } + {\bf n}\ .\ \dot {\bf
G}_1\right) - \frac 1{r^2} \left(|f| + {\bf n}\ .\ {\bf G}_1 +
{\bf n}\ .\ \dot {\bf G}\right)$$ around $r_{\cal M} = t -
\tau_{\cal M}$, where $\tau_{\cal M}$ is the value of $\tau$ at
which the function $|f|$ has maximum. In fact, we can put $r =
r_{\cal M}$ in the argument of all functions except for $|f|\dot{\
}$, which we must expand up to the first order in $\Delta r = r -
r_{\cal M}$. After replacing $r_{\cal M}$ by $t$ we obtain
\begin{equation} \Delta r_M^{(0)} = \frac 1{|f|_{\cal M} \mbox{\hskip -2.5mm
\large $\ddot{}$} \hskip 2.5mm} \left[{\bf n}\ .\ {\dot {\bf
G}}_{1{\cal M}} + \frac 1t \left(|f|_{\cal M} + {\bf n}\ .\ {\bf
G}_{1{\cal M}} + {\bf n}\ .\ \dot {\bf G}_{\cal M} \right)
\right], \label{eq:Dr0}
\end{equation}
where the index ${\cal M}$ indicates that the quantity in question
is to be taken at $r = r_{\cal M}$ (or, equivalently, at $\tau =
\tau_{\cal M}$). By differentiating this with respect to $t$ we
finally find that the correction to the group velocity coming from
the nonplanarity of the wave packet is
\begin{equation} \Delta v^{(0)} =
- \frac 1{|f|_{\cal M} \mbox{\hskip -2.5mm \large $\ddot{}$}
\hskip 2.5mm t^2} \left(|f|_{\cal M} + {\bf n}\ .\ {\bf G}_{1{\cal
M}} + {\bf n}\ .\ \dot {\bf G}_{\cal M} \right). \label{eq:Dv0}
\end{equation}
To estimate the dominant term, note that $|f|_{\cal M}
\mbox{\hskip -2.5mm \large $\ddot{}$} \hskip 2.5mm \sim -
|f|_{\cal M}/\sigma_r^2$, where $\sigma_r$ is the dispersion of
the packet in the radial coordinate. This yields
$$\Delta v^{(0)} \sim \frac {\sigma_r^2}{t^2}.$$
We can see that the group velocity of the radiation emitted by a
compact source is superluminal even without a time-dependent
medium, merely due to the fact that the factor $1/r$ shifts the
maximum of the envelope of the packet the more backwards the
closer the packet to the source. This is demonstrated in Fig.
\ref{fig:spp}.

\vskip 2mm If we put the field $V$ back into the wave equation, we
obtain, in addition to $\Delta v^{(0)}$, another contribution to
the group velocity $\Delta v_V$. As we shall see, nonplanarity has
significant effect on the propagation of the packet, but in
relation to $\Delta v_V$ we can make use of the fact that in a
sufficiently small domain any spherical wave can be replaced by
planar, and identify $\Delta v_V$ with the contribution to the
group velocity $\Delta v$ computed before. To estimate the
relative size of the corrections $\Delta v_V$ and $\Delta
v^{(0)}$, consider again, as in section \ref{sec:vg}, a dust
universe. Then $\Delta v_V \sim (kt)^{-2}$, and since the
dispersion in $x$-space $\sigma_r$ equals approximately the
inverse of the dispersion in $k$-space $\sigma$, we have
\begin{equation}
\frac {\Delta v_V}{\Delta v^{(0)}} \sim \frac {\sigma^2}{k^2} \ll
1. \label{eq:ratio}
\end{equation}
From (\ref{eq:Dv0}) we can estimate the next-to-leading term in
$\Delta v^{(0)}$ as $\Delta v^{(0)}_{next} \sim v_s \Delta
v^{(0)}$, where $v_s$ is the typical velocity with which the parts
of the source move with respect to each other. This yields
\begin{equation}
\frac {\Delta v_V}{\Delta v^{(0)}_{next}} \sim \frac
{\sigma^2}{v_s k^2}. \label{eq:ratio1}
\end{equation}
The leading term in $\Delta v^{(0)}$ is determined by the form of
the packet in the vicinity of the peak of the envelope, and can in
principle be established by the local measurement of the
oscillating field. Thus, $\Delta v_V$ can be found from local data
if it dominates $\Delta v^{(0)}_{next}$, which is the case if the
velocity of the internal motion of the source is small enough,
$v_s \ll \sigma^2/k^2$.


\section{Conclusion}
\label{sec:concl}

When inspecting the equation for gravitational waves in an
expanding universe, one immediately observes that the group
velocity predicted by it is superluminal provided the pressure to
density ratio is less than 1/3 (which is surely the case in our
universe) and the typical wavelength of the waves lays well within
the horizon. The positive sign of the additional velocity $\Delta
v$ is entirely the consequence of the negative sign of the
'external field' $V$, which in turn reflects the fact that for
gravitational waves there exists a finite Jeans length of order of
horizon length. Note that for electromagnetic waves, $\Delta v$ is
exactly zero. The reason is the conformal invariance of Maxwell
action in four dimensions, which prevents the intensity of the
field from having 'tails' inside the light cone; for details, see
\cite{sofa}. To support our calculation of $\Delta v$, we have
performed a detailed analysis of wave propagation in a
time-dependent medium. The value of $\Delta v$ proved correct for
plane waves, and one would expect that it will be automatically
correct for spherical waves, too, since group velocity is a local
characteristic and spherical waves are locally equivalent to plane
ones. However, the effect is so small that the effects of
nonplanarity turn out to be substantially superior to it. For
gravitational waves coming from astrophysical sources, the net
time shift $\Delta \tau$ caused by $\Delta v$ is at best of order
of 10 ps. Such a short time would be surely hard to measure even
if its beginning and end were defined by single events, which they
are not. To determine $\Delta \tau$, one should reconstruct the
envelope of the pulse in the vicinity of the peak with an error
less by many orders of magnitude than the period of the pulse;
calculate the position of the peak relative to the forefront at
the moment of the formation of the pulse with the same accuracy;
and separate out the effects of nonplanarity. A principal obstacle
to the first step seems to be that in order to carry it out, one
needs to know the imaginary part of the oscillating field, which
is an auxiliary object inaccessible to the measurements. However,
the analysis in the appendix suggests that the complex field can
be reconstructed from the data by a suitable approximate
procedure, most simply, by determining the Fourier coefficients of
the real field seen in the experiment and making use of the
coefficients with positive $k$'s only. While this would typically
produce a much greater error than the measured effect, only the
time-independent part of the position of the peak of the envelope
(the quantity $s_0$ in the notations of the appendix) would be
affected. Obviously, there is no hope that the outlined procedure
to measure $\Delta \tau$ could sometimes be carried out in
practice. Nevertheless, we believe that the effect is still of
some interest since it contributes to our understanding of how the
propagation of gravitational waves is affected by the expansion of
the universe.


\vskip 4mm {\it Acknowledgement.} This work was supported by the
grant VEGA 1/3042/06.


\appendix
\section{Propagation of planar wave packets}
\label{ap:planar} \setcounter{equation}{0}
\renewcommand{\theequation}{A-\arabic{equation}}
A general planar wave packet in a time-dependent medium is
described by the integral
\begin{equation} {\cal U} = \int \limits_{-\infty}^\infty a(K)
e^{-i \Delta \psi + i\Delta Kx} dK,\label{eq:scrU}
\end{equation}
where
\begin{equation} \Delta \psi = \psi(K, t) - \psi = \xi \Delta K +
\frac 12 \eta \Delta K^2 + \frac 16 \zeta \Delta K^3 + \ldots, \ \
\Delta K = K - k.\label{eq:dpsi}
\end{equation}
Introduce the Fourier transform of $a$,
$$\alpha (s) = \int \limits_{-\infty}^\infty a(K) e^{i\Delta Ks}
dK.$$ If we write the exponential function in the expression for
$\cal U$ as
$$e^{-i \Delta \psi + i\Delta Kx} = e^{i\Delta K {\cal X}}
\left(1 - \frac i2 \eta \Delta K^2 - \frac i6 \zeta \Delta K^3 +
\ldots \right),$$ where ${\cal X} = x - \xi$, we find
$${\cal U} = \alpha + \frac i2
\eta \alpha'' + \frac 16 \zeta \alpha''' + \ldots,$$ where the
functions $\alpha$, $\alpha''$, $\ldots$ are taken at the
(complex) value of the argument $s = \cal X$. Next, we expand the
functions $\alpha$, $\alpha''$, $\ldots$ around the point $s_0$ at
which $|\alpha|$ has maximum. This yields
\begin{equation} {\cal U} = A_0 + A_1 \Delta {\cal X} +
\frac 12 A_2 (i\eta + \Delta {\cal X}^2) + \frac 16 A_3 (\zeta +
3i \eta \Delta {\cal X} + \Delta {\cal X}^3) + \ldots,
\label{eq:expU}
\end{equation}
where $A_n$ is the $n$th derivative of $\alpha$ at $s = s_0$ and
$\Delta {\cal X} = {\cal X} - s_0$. Suppose $a(K)$ is of the form
$f(\Delta K/\sigma)/\sigma$, where $|f|$ has maximum of order 1,
$f(u)$ is significantly different from zero only in the interval
$|u|\lesssim 1$, and the derivatives of $f$ obey $|f^{(n)} (u)|
\lesssim 1$ everywhere in the interval $|u|\lesssim 1$. For a
large class of such functions (for example, for Gaussian functions
with the dispersion of order $\sigma$ multiplied by appropriately
normalized, but otherwise arbitrary polynomials), $\alpha (s)$ is
of the form $\tilde f(\sigma s)$, where $\tilde f(u)$ has the same
properties as $f(u)$. Then, the coefficients in the sum
(\ref{eq:expU}) can be estimated as
$$A_n = \tilde f^{(n)}(\sigma s_0)\sigma^n \sim \sigma^n,$$
and the sum is in fact a power series in $\sigma$. The real part
of the shift $\Delta {\cal X}$ is $\Delta X = X - s_0$, where $X =
{\cal X}_R = x - \xi_R$. Here, as in section \ref{sec:vg1}, we
denote by $R$ and $I$ the real and imaginary part of the quantity
in question. If we pass from $\Delta {\cal X}$ to $\Delta X$ in
the expression for $\cal U$, we obtain a series
\begin{equation} {\cal U} = u_0 + u_1 \Delta X + u_2 \Delta X^2
+ \ldots, \label{eq:expU1}
\end{equation}
with each coefficient itself a series,
\begin{equation} u_n = \frac 1{n!} \sum \limits_{m = 0}^\infty
A_{n + m}\psi_m,\ \ \psi_0 = 1,\ \ \psi_1 = - i\xi_I,\ \ \psi_2 =
\frac 12 (i\eta - \xi_I^2),\ \ \psi_3 = \frac 16 (\zeta + 3\xi_I
\eta + i\xi_I^3),\ \ \ldots \label{eq:uk}
\end{equation}
We are interested in the shift of the peak of the wave packet $X_M
= x_M - \xi_R$. This can be written as $X_M = s_0 + \Delta X_M$,
where $\Delta X_M$ is the value of $\Delta X$ at which the
function $|{\cal U}|^2$ has maximum at fixed $t$. Extrema of
$|{\cal U}|^2$ are zeros of the function
$$\partial_x |{\cal U}|^2 \propto U_0 + U_1 \Delta X + U_2 \Delta X^2 +
\ldots,\ \ U_n = \sum \limits_{m = 0}^n (n + 1 - m) (u_m u_{n + 1
- m}^* )_R.$$ For small enough $\sigma$ we expect $|{\cal U}|$ to
be close to $|\alpha|$ at $X \sim s_0$. Then, the value of $\Delta
X_M$ can be found by solving the equation $U_0 + U_1 \Delta X +
U_2 \Delta X^2 + \ldots = 0$ iteratively,
$$\Delta X_M^{(1)} = - \frac {U_0}{U_1} \equiv \Delta,\ \ \Delta
X_M^{(2)} = \Delta - \frac {U_2}{U_1} \Delta^2,\ \ \Delta
X_M^{(3)} = \Delta - \frac {U_2}{U_1} \Delta^2 - \left[\frac
{U_3}{U_1} - 2 \left( \frac {U_2}{U_1}\right)^2\right] \Delta^3,\
\ \ldots$$ The quantities $U_k$ are generically of order
$\sigma^{k + 1}$, except for $U_0$ which is of order $\sigma^2$.
Indeed,
$$U_0 = (u_0 u_1^* )_R = (A_0 A_1^* )_R +
O(\sigma^2),$$ and if we fix, for the sake of simplicity, the
overall phase of $\alpha$ so that $\alpha(s_0)$ is real, we find
that $A_0$ is real and $A_1$ purely imaginary. From the estimate
of $U_k$ it follows that the term proportional to $\Delta^n$ in
the iterative expression for $\Delta X_M$ is generically of order
$\sigma^{n - 1}$, so that the procedure is meaningful for small
enough $\sigma$. Finally, we expand all terms entering $\Delta
X_M$ in the powers of $\sigma$ to obtain
\begin{equation} \Delta X_M = X_{M0} + X_{M1} + X_{M2} + \ldots,\ \
X_{Mn} \sim \sigma^n. \label{eq:dxgen}
\end{equation}
Note that $s_0$ is generically of order $1/\sigma$ (the width of
the wave packet in $x$-space), therefore a power series for
$\Delta X_M$ defines a power series for $X_M$, too, but starting
with a term of order minus one rather than zero. If we write
$$U_0 = {\cal A} + \Delta {\cal A} + \ldots,\ \ U_1 = {\cal
B} + \Delta {\cal B} + \ldots,\ \ U_3 = {\cal C} + \ldots,$$ where
$\cal A$, $\cal B$ are of order $\sigma^2$ and $\Delta \cal A$,
$\Delta \cal B$, $\cal C$ are of order $\sigma^3$, we find that
the first two terms in the expansion of $\Delta X_M$ are
\begin{equation} X_{M0} =  - \frac {\cal A}{\cal B} \equiv \delta,\ \
X_{M1} = - \frac 1{\cal B} (\Delta {\cal A}+ \Delta {\cal B}
\delta + {\cal C} \delta^2). \label{eq:d01}
\end{equation}
(We do not consider the case ${\cal B} = 0$ which requires a
separate discussion.) The expressions for $\Delta \cal A$, $\Delta
\cal B$ and $\cal C$ are obtained from the expansions of first few
$u$'s in the powers of $\sigma$. A straightforward computation
yields
\begin{eqnarray}
\lefteqn{{\cal A} = A_0 A_{2I}\xi_I,\ \ {\cal B} = A_0 A_{2R} +
A_{1I}^2,\ \ \Delta {\cal A} = - \frac 12 A_0 [A_{3R} (\eta_I +
\xi_I^2) + A_{3I} \eta_R] + \frac 12 A_{1I} \mbox{\tiny $\times$}
\nonumber} \\
& \mbox{\tiny $\times$} [A_{2R}\eta_R - A_{2I} (\eta_I -
\xi_I^2)],\ \ \Delta {\cal B} = (A_0 A_{3I} - A_{1I} A_{2R})
\xi_I,\ \ {\cal C} = \frac 12 (A_0 A_{3R} + 3A_{1I} A_{2I}).
\label{eq:ABC}
\end{eqnarray}
Note that only $\xi$ and $\eta$ (the first and second derivative
of $\psi$) appear here, although $\zeta$ (the third derivative) is
to be included, too, into the formulas for $\cal U$ and $|{\cal
U}|^2$, if one writes them with the accuracy needed for the
calculation of $\Delta X_M$ up to the term $X_{M1}$. The point is
that $\zeta$ enters only the term independent on $\Delta X$ in
$|{\cal U}|^2$ and therefore plays no role in the extremization.
Note, furthermore, that if $A_{2I}$ is zero (the phase of the
function $\alpha$ has zero second derivative at the maximum of
$|\alpha|$), $X_{M0}$ is zero and $\Delta X_M$ is at least of
order $\sigma$. For this class of packets, $\Delta X_M$ can be
made arbitrarily small by shrinking the packet in $k$-space. To
understand why $A_{2I}$ rather than $A_{1I}$ determines $X_{M0}$,
note that $A_{1I}$ can be adjusted arbitrarily by shifting the
packet in $k$-space. But the shift of the packet by the distance
$q$ results only in multiplication of ${\cal U}_{lin} =
\alpha({\cal X})$ (the function ${\cal U}$ in the linear
approximation) by the factor $e^{iq\cal X}$, which has no effect
on the position of the peak of $|{\cal U}_{lin}|$ at fixed $t$.
Thus, we can change $A_{1I}$ without changing $X_M$ in the linear
approximation; and since the linear approximation is all what is
needed to determine $X_{M0}$, the value of $X_{M0}$ cannot depend
on $A_{1I}$.

\vskip 2mm An important special case is the Gaussian wave packet
(\ref{eq:gauss}). The Fourier transform of a Gaussian amplitude is
again Gaussian,
$$\alpha_G = \sqrt{2\pi}\sigma \exp \left(-\frac 12 \sigma^2
s^2\right).$$ Since this is real, all imaginary $A$'s vanish; and
since this is an even function of $s$, with the maximum at the
origin, $s_0 = 0$ and all odd $A$'s vanish, too. This implies that
$X_{M0} = X_{M1} = 0$ and the leading term in $\Delta X_M$ as well
as in $X_M$ is $X_{M2}$. For an arbitrary packet with real nonzero
$A_0, A_2, A_4, \ldots$ and zero $A_1, A_3, \ldots$ we have
$$U_0 = (u_0 u_1^*)_R = \frac 16 A_0 A_4 (\zeta_R + 3\xi_I \eta_R)
+ \frac 12 A_2^2 \xi_I \eta_R + \ldots,\ \ U_1 = 2(u_0 u_2^*)_R +
|u_1|^2 = A_0 A_2 + \ldots$$ If we insert here the $A$'s for a
Gaussian packet, $A_{2\nu} = (-1)^\nu (2\nu - 1)!! \sigma^{2\nu}$,
and use the resulting $U_0$ and $U_1$ in the formula $X_M =
-U_0/U_1 + \ldots$, we find
\begin{equation} X_{M2} = \left(\xi_I \eta_R + \frac 12 \zeta_R
\right)\sigma^2. \label{eq:dxgauss}
\end{equation}
Note that the first term can be obtained also from the analytical
expression for $\cal U$ in the quadratic approximation, with the
$\Delta K$-expansion of $\psi$ truncated after the $\eta$-term. In
this approximation, $\cal U$ can be found from the formula for
${\cal U}_{lin}$ written down in section \ref{sec:vg1}, by
replacing
$$\frac 1{\sigma^2} \to \frac 1{\sigma^2} + i\eta.$$
This yields
$$|{\cal U}| = \frac 1{\sqrt A} e^{- \frac
1{2A^2} \sigma^2 [a(X^2 - \xi_I^2) - 2\xi_I \eta_R \sigma^2 X]},$$
where $a = 1 - \eta_I \sigma^2$ and $A = \sqrt{a^2 + \eta_R^2
\sigma^4}$. The wave packet is Gaussian as in the linear
approximation, but with a time-dependent width $\sigma_x =
A/(\sqrt{a} \sigma)$ and the peak shifted by $X_M = \xi_I \eta_R
\sigma^2/a$. Within the current accuracy, the factor $1/a$ in the
expression for $X_M$ plays no role. Thus, $X_M$ effectively equals
the first term on the right hand side of (\ref{eq:dxgauss}) as
mentioned above.

\vskip 2mm Let us find conditions under which our perturbative
approach is applicable, and estimate the time-dependent part of
$X_M$. We are interested in this part only, because we wish to
compare different contributions to $v = \dot x_M$ rather than to
$x_M$. If we use the ``reduced'' definition of $\psi$, with
whatever factor in $\omega$ depending only on $k$ removed from the
term proportional to $\log \omega$, the time-dependent part of
$X_M$ is just $\Delta X_M$. (If we left such factor in $\omega$,
we would have to skip from $X_M$, in addition to $s_0$, also a
part of $X_{M0}$.) As demonstrated in previous calculations, the
term $X_{Mn}$ in the expansion of $\Delta X_M$ is a sum of
products
$$\xi_I^{\nu_{1I}} \eta_R^{\nu_{2R}} \eta_I^{\nu_{2I}} \zeta_R^{\nu_{3R}}
\zeta_I^{\nu_{3I}} \ldots \psi_R^{(n + 1)\nu_{n + 1,R}} \psi_I^{(n
+ 1)\nu_{n + 1,I}} \sigma^n,\ \ \sum \limits_{mA} m\nu_{mA} = n +
1,$$ multiplied by coefficients of order 1. For the real and
imaginary part of the $n$th derivative of $\psi$ we have
$\psi_R^{(n)} \sim \hat \eta_R/\kappa^n$ and $\psi_I^{(n)} \sim
\hat \xi_I/\kappa^n$, where $\hat \xi = \xi \kappa$, $\hat \eta =
\eta \kappa^2$ and $\kappa$ is the scale of $k$ on which $\omega$
varies. Thus, if $p_{(\nu_R, \nu_I)}$ is the sum of terms in
$X_{Mn}$ with given $\nu_R = \sum \nu_{mR}$ and $\nu_I = \sum
\nu_{mI}$ such that $n' = 2\nu_R + \nu_I \le n + 1$, we have
\begin{equation} p_{(\nu_R, \nu_I)} \sim \hat \xi_I^{\nu_I}
\hat \eta_R^{\nu_R} \frac {\sigma^n} {\kappa^{n + 1}}.
\label{eq:s2estim}
\end{equation}
Rewrite this as
$$p_{(\nu_R, \nu_I)} \sim \bigg[\xi_I (\xi_I \sigma)^{\nu_I - 1}
(\eta_R \sigma^2)^{\nu_R}\ \ \mbox{or}\ \ \eta_R \sigma (\xi_I
\sigma)^{\nu_I} (\eta_R \sigma^2)^{\nu_R - 1}\bigg] \left(\frac
\sigma\kappa \right)^{n + 1 - n'},$$ where the first expression is
to be used if $\nu_R = 0$, the second expression is to be used if
$\nu_I = 0$, and both expressions are equally good if $\nu_R \ne
0$ as well as $\nu_I \ne 0$. As can be seen from this estimate, we
must require
\begin{equation} |\xi_I| \sigma \ll 1,\ \ |\eta_R| \sigma^2 \ll
1,\ \ \frac \sigma \kappa \ll 1,  \label{eq:ptcond}
\end{equation}
in order that the parameter $\sigma$ can be considered small. This
guarantees that all $p_{(\nu_R, \nu_I)}$, as well as their sum
$X_{Mn}$, are suppressed the more the higher the value of $n$. In
particular, if all three parameters in (\ref {eq:ptcond}) are of
the same order $\epsilon$, $p_{(\nu_R, \nu_I)}$ and $X_{Mn}$ are
of order $\epsilon^n$. In the perturbation theory we can estimate
the quantity of interest by the leading term, or terms, of the
expansion in the small parameter of the theory. For a general
packet, the estimate of $p_{(\nu_R, \nu_I)}$ as well as the
explicit expressions for $X_{M0}$ and $X_{M1}$ suggest that the
leading terms in $\Delta X_M$ are
\begin{equation} \Delta X_M \sim \xi_I + \eta_R \sigma.
\label{eq:genest}
\end{equation}
For a Gaussian packet, $X_M$ in the lowest nonzero order in
$\sigma$ is
\begin{equation}
X_{M2} \sim \left(\xi_I + \frac 1\kappa \right) \eta_R \sigma^2.
\label{eq:dxgauss1}
\end{equation}
Any term in $X_M$ that {\it cannot} be obtained from these two by
multiplying them by some product of powers of the three parameters
in (\ref {eq:ptcond}), is of the form $\xi_I (\xi_I \sigma)^n
(\sigma/\kappa)^m$ or $\eta_R \sigma (\eta_R \sigma^2)^n$ (is
proportional either to a product of some powers of $\xi_I$ and its
derivatives, or to some power of $\eta_R$). We will argue that
such terms are actually missing, so that the two terms of the
lowest nonzero order in $\sigma$ are at the same time the leading
terms. The $n$th term in the iterative expression for $X_M$ is
proportional to $\Delta^n$, that is, to $U_0^n$. If we write $U_0$
as
$$U_0 = \sum \limits_{mn} A_m A_{n + 1}
(\psi_m \psi_n^*)_R \sim \sum \limits_{\nu} \sigma^{2\nu} \sum
\limits_{\mu \le \nu} (\psi_{2\mu} \psi_{2\nu - 2\mu - 1})_R,$$ we
can see that in order that there exists a term in $X_M$ that
assumes one of the two forms mentioned above, there must exist at
least one expression of the type $(\psi_{even} \psi_{odd})_R$
which assumes the same form with skipped $\sigma$ (since any term
in $X_M$ contains at least one such expression). The $\psi$'s can
be written as
$$\psi_n \sim i^n (\Psi_n + \xi_I \Psi_{n - 1} + \ldots +
\xi_I^{n - 2}\Psi_2 + \xi_I^n),$$ where
$$\Psi_n \sim i\psi^{(n)} + \psi^{(n_1)} \psi^{(n_2)} + i \psi^{(n_1)}
\psi^{(n_2)} \psi^{(n_3)} + \ldots,\ \ n_i \ge 2,\ \ \sum n_i =
n.$$ The terms in $\psi_n$ containing only $\xi_I$ and its
derivatives are of the form $i^n \xi_I^m \psi_I^{(n_1)}
\psi_I^{(n_2)} \ldots$, with no factor $i$ contributed by
$\Psi_m$, thus the terms of the same kind in $\psi_{even}
\psi_{odd}$ are purely imaginary and none of them appears in
$(\psi_{even} \psi_{odd})_R$. The quantity $\psi_n$ does not
include terms containing just $\eta_R$ if $n$ is odd, since there
is necessarily at least one odd derivative of $\psi$ present in
every contribution to $\Psi_n$ in this case, therefore the
expression $\psi_{even} \psi_{odd}$ does not include such terms
either. Thus, there are no terms of the type $\xi_I^n/\kappa^m$ or
$\eta_R^n$ in $(\psi_{even} \psi_{odd})_R$, q. e. d. As a result,
the estimate of $X_M$ coincides with the estimate of $X_{M2}$.
However, one of its two terms can be omitted if we note that
$|\xi_I| \lesssim 1/\kappa$. (For a general dispersion relation we
have $\xi_I \sim 1/\kappa$, and for special dispersion relations
we can have, when using the ``reduced'' definition of $\psi$, also
$|\xi_I| \ll 1/\kappa$.) This yields
\begin{equation}
X_M \sim \frac 1\kappa \eta_R \sigma^2. \label{eq:gaussest}
\end{equation}


\end{document}